\documentclass[aps,pra,twocolumn,showpacs,floatfix]{revtex4}

\usepackage{dcolumn}
\usepackage{graphicx}
\usepackage{epsf}
\usepackage{amsmath}
\usepackage{bm}
\usepackage{times}

\newcommand{\addrMST}{
Department of Physics, Missouri University of Science and Technology,
Rolla, Missouri, MO65409-0640, USA}

\begin{document}

\sloppy

\title{Binding two-loop vacuum-polarization corrections to the bound-electron 
$\bm g$ factor}

\author{Ulrich D.~Jentschura}
\affiliation{\addrMST}

\begin{abstract}
We commence the evaluation of the one- and two-loop binding corrections to the
$g$ factor for an electron in a hydrogenlike system of order ${{\alpha^2
(Z\alpha)^5}}$ and consider diagrams with closed fermion loops.  The one-loop
vacuum-polarization correction is rederived and confirmed.  For the two-loop
vacuum-polarization correction, due to a specific gauge-invariant set of
diagrams with closed fermion loops, we find a correction 
$\delta g =  7.442 \, (\alpha/\pi)^2 \, (Z\alpha)^5$.  
Based on the numerical trend of the coefficients
inferred from the gauge-invariant subset, we obtain a numerically large
tentative estimate for the complete two-loop binding correction to the $g$
factor (sum of self-energy and vacuum polarization).
\end{abstract}


\maketitle

%
%
\section{INTRODUCTION}
\label{intro}

The bound-electron $g$ factor has been the subject of intense investigations
over the past decade, both experimentally as well as theoretically.  It
describes the response of the bound electron to an external homogeneous
magnetic field and is naturally different from the $g$ factor of a free
electron, due to the binding of the electron to the nucleus.  Recent
measurements for hydrogenlike ions with a spinless nucleus in the region of low
nuclear charge number $Z$ have been reported and discussed in
Refs.~\cite{Be2000,HaEtAl2000prl,BeEtAl2002prl,VeEtAl2004}.

For precision experiments with trapped hydrogenlike ions, the
most important atomic state to be considered is the ground state, and we
restrict the discussion to the $g$ factor of the electronic ground state, for
which we write $g \equiv g(1{\rm S})$.  From the relativistic (Dirac) theory of
the bound electron (which does not include radiative corrections), one
obtains~\cite{Br1928}
\begin{equation}
g = 2 - \frac23\, (Z\alpha)^2 - 
\frac16\, (Z\alpha)^4 + {\cal O}{(Z\alpha)^6} \,.
\end{equation}
Here, $\alpha$ is the fine-structure constant, and 
$Z$ is the nuclear charge number.
The negative sign of the correction terms of higher order in the 
$Z\alpha$-expansion imply that $g < 2$ for higher nuclear charge 
numbers $Z$. Therefore, planned experiments in the 
high-$Z$ region~\cite{ShEtAl2006} have been termed ``2--$g$''-experiments. 

The quantum electrodynamic (QED) corrections to the bound-electron $g$ factor
can be expressed as a combined expansion in $\alpha$ and $Z\alpha$, where the
latter parameter describes the strength of the coupling to the
nucleus~\cite{JeEv2005}.  The first few terms in the expansion of the 
one-loop correction $\delta g^{(1)}$ to the bound-electron $g$ factor
(sum of self energy and vacuum polarization) 
in powers of $Z\alpha$ read~\cite{Ka2000,PaJeYe2004} 
\begin{align}
\delta g^{(1)} =& \; \frac{\alpha}{\pi} \, \biggl\{ 
1 + \frac{(Z\alpha)^2}{6} 
+ (Z\alpha)^4 \biggl[ \frac{32}{9}\, {\ln}[(Z\alpha)^{-2}] \\[2ex]
& \; - 11.303\,191 \biggr] 
+ a_{50} \, (Z\alpha)^5 + {\cal O}(Z\alpha)^6 \biggr\}\,.
\nonumber
\end{align}
According to commonly accepted conventions,
the coefficient $a_{50}$ carries two indices, the first of which 
counts the power of $Z\alpha$, whereas the second counts the 
power of the logarithm ${\ln}[(Z\alpha)^{-2}]$. 

The two-loop correction reads~\cite{PaCzJeYe2005}
\begin{align}
& \delta g^{(2)} = \;
\left( \frac{\alpha}{\pi} \right)^2\, \biggl\{ 
{-0.656\,958} \,  {\bigg(} 1 + 
\frac{(Z\alpha)^2}{6} {\bigg)}  
+ (Z\alpha)^4 \, \\[2ex]
& 
\times \biggl\{ \frac{56}{9} {\ln}[(Z\alpha)^{-2}]  
- 16.436\,842 
+ b_{50} (Z\alpha)^5 + {\cal O}(Z\alpha)^6 \biggr\}.
\nonumber
\end{align}
Our goal here is to evaluate the contribution to $b_{50}$ due to a subset of
the diagrams containing closed fermion loops, and to rederive the known result
for the vacuum-polarization contribution to $a_{50}$.  We recall that according
to Fig.~21 of Ref.~\cite{Be2000}, the number of two-loop diagrams contributing
to the $g$ factor is large, and the particular diagrams considered here form
one of the most straightforward gauge-invariant subsets in two-loop order.  As
the whole set of two-loop diagrams can be broken up into smaller
gauge-invariant subsets, the evaluation could be initiated by considering
gauge-invariant subsets. Since the diagrams for the $g$ factor are related to
those for the Lamb shift (except for the additional presence of an external
magnetic field), a byproduct of our calculations is a confirmation of results
obtained previously for the contribution of corresponding diagrams to the
two-loop, binding correction to the Lamb shift~\cite{Pa1993pra} of order
$\alpha^2 (Z\alpha)^5$.

This brief communication is organized as follows.
After a discussion of the one-loop correction 
in the order $\alpha\, (Z \alpha)^5$ (see Sec.~\ref{oneloop}),
we describe the two-loop calculations in the order
$\alpha^2\, (Z \alpha)^5$ in Sec.~\ref{twoloop}.
Conclusions are drawn in Sec.~\ref{conclusions}.

%
%
\section{ONE--LOOP CORRECTION}
\label{oneloop}

First, we would like to rederive the leading vacuum-polarization correction to
the bound-electron $g$ factor of order $\alpha(Z\alpha)^4$. To this end, we
recall that for the interaction of an electron with a constant magnetic field,
one can derive the following, effective Hamiltonian based on long-wavelength
quantum electrodynamics~\cite{Pa2004} for the interaction of an electron with
an external static magnetic field $\vec{B}$,
\begin{align}
H_\sigma = & \;
e \, \vec{\sigma} \cdot \vec{B} \left(
- \frac{1}{2 m} 
+ \frac{\vec{p}^{\,2}}{4 m^3} 
-\frac{1}{12 \, m^2} \, 
(\vec{r} \cdot \vec\nabla V) \right) \,,
\label{HsigmaRRR}
\end{align}
where $\vec{p}$ is the bound-electron momentum,
$m$ is the electron mass, and
$V$ is the total static potential felt by the electron.
This potential can be either the Coulomb potential, which we denote by
$V_C$ in the following,
or a vacuum-polarization correction $\delta V$.

We now briefly recall how to evaluate of the 
one-loop vacuum-polarization correction based on 
the effective Hamiltonian \eqref{HsigmaRRR} 
and on well-known formulas for vacuum-polarization effects.
Indeed, we use the well-known Uehling approximation 
for the vacuum-polarization potential
and identify the potential in \eqref{HsigmaRRR} 
as $V \to \delta V \to V_U$,
\begin{equation}
\label{uehl}
V_U({\vec r}) = \frac{\alpha}{\pi}\int_0^1
dv\,\frac{v^2(1-v^2/3)}{1-v^2}
\,\exp\left(-\lambda\,r\right)\,
\left[\frac{-Z \alpha}{r}\right] \,,
\end{equation}
with $\lambda = 2\,m/\sqrt{1-v^2}$.

The  first correction $E_1$ to the spin-dependent magnetic-field 
interaction energy (and thus to the $g$ factor)
is obtained if we replace $V \to V_U$ in the third
term in brackets in Eq.~\eqref{HsigmaRRR},
\begin{align}
\label{E1}
E_1 =& \; 
\left< \phi \left|
- \frac{e}{12\,m^2}\,
(\vec{r} \cdot \vec{\nabla} V_U)\,(\vec{\sigma}\cdot\vec{B})
\right| \phi \right>
\nonumber\\[2ex]
= & \; \frac13 \, \left< \phi \left|
\vec{r} \cdot \frac{\vec{\nabla} V_U}{m} \,
\right| \phi \right>
\left< -\frac{e}{4 \, m} \,\vec{\sigma}\cdot\vec{B} \right> \,,
\end{align}
where $| \phi \rangle$ denotes the nonrelativistic 
atomic ket vector corresponding to the atomic state 
under investigation (here, the
ground state). Of course, the rightmost term in Eq.~\eqref{E1} is 
evaluated on the bound-state wave function, but we write it as being
proportional to $\left< \vec{\sigma}\cdot\vec{B} \right>$, where it is
understood that for a $S$ state, the spin is either pointing up or down.  This
means that the expectation value on the right-hand side is to be evaluated
using the spin degrees of freedom only, and it is therefore denoted by a simple
bracket. Because $E_1$ is a first-order spin-dependent 
energy correction in a uniform external magnetic field,
it can be related directly to a correction to the $g$ factor.
For this purpose, we write the interactions as multiplicative
corrections to the normalized interaction 
$-\frac{e}{4 \, m} \,\vec{\sigma}\cdot\vec{B}$;
the latter leads to a $g$ factor of unity.

The correction $E_1$, which is a first-order correction, 
now has to be supplemented by some second-order effects.
Let us therefore consider the case where $V$ in 
Eq.~\eqref{HsigmaRRR} represents the Coulomb potential $V_C$.
In order to evaluate the second-order effects, we investigate
the Uehling correction
in conjunction with the second and the third term
in round brackets in Eq.~\eqref{HsigmaRRR}, which represent 
corrections to the $\vec{\sigma}\cdot \vec{B}$ interaction of 
relative order $(Z\alpha)^2$. The perturbation to the wave
function induced by the leading-order interaction
$-e \vec\sigma \cdot\vec B/(2m) = 
-g\, e \vec\sigma \cdot\vec B/(4m)$ vanishes. 

The first of the nonvanishing second-order 
effects is obtained by considering a second-order 
perturbation involving the Uehling potential and the 
second term in brackets in \eqref{HsigmaRRR}
\begin{align}
E_2 =& \; 2 \, 
\left< \phi \left| V_U \, \left( \frac{1}{E - H} \right)' \, 
\left( \frac{\vec{p}^2}{4 m^3} \, e \,\vec{\sigma}\cdot\vec{B}
\right) \right| \phi \right> 
\nonumber\\[2ex]
=& \; 4 \, \left< \phi \left| \frac{V_U}{m} \, 
\left( \frac{1}{E - H} \right)' \, V \right| \phi \right>
\left< -\frac{e}{4 \, m} \,\vec{\sigma}\cdot\vec{B} \right>
\end{align}
The second of these is obtained by considering again the 
third term in brackets in \eqref{HsigmaRRR}, but this time 
acting on the Coulomb potential $V$ in second-order perturbation theory,
\begin{align}
E_3 =& 2\,
\left< \phi \left| \frac{V_U}{m} \left( \frac{1}{E - H} \right)' 
\left( - \frac{e}{12\,m} [(r \cdot \vec{\nabla}) V_C] 
\vec{\sigma}\cdot\vec{B}
\right) \right| \phi \right> 
\nonumber\\[2ex]
=& -\frac{2}{3} \,
\left< \phi \left| \frac{V_U}{m} \, \left( \frac{1}{E - H} \right)' \,
V_C \right| \phi \right>
\left< -\frac{e}{4 \, m} \,\vec{\sigma}\cdot\vec{B} \right>
\end{align}
Taking into account the Hellmann--Feynman theorem,
\begin{equation}
\left( \frac{1}{E - H} \right)' \,
\left. \left. V_C \right| \phi \right> = 
\left. \left.  Z \frac{\partial}{\partial Z} \right| \phi \right>\,,
\end{equation}
the sum of the corrections 
$E_1 + E_2 + E_3$
leads to the known result~\cite{Ka2000}
\begin{equation}
\label{deltag}
\delta g = 
\frac13 \, \left< \phi \left|
\vec{r} \cdot \frac{\vec{\nabla} V_U}{m} \,
\right| \phi \right> +
\frac{10}{3} \,
\left< \phi \left| \frac{V_U}{m} \,
Z \frac{\partial}{\partial Z} \right| \phi \right> \,.
\end{equation}
In the lowest-order in the $Z\alpha$ expansion, 
Eq.~\eqref{uehl} then immediately leads to the leading-order 
vacuum-polarization correction to the $g$ factor~\cite{Ka2000},
\begin{equation}
\label{gU}
\delta g^{(1)}_{U} = \frac{\alpha}{\pi} \, (Z\alpha)^4 \,
\left( -\frac{16}{15} \right)\,,
\end{equation}
where the index $U$ reminds us of the Uehling potential.

We now consider the 
wave function slope and the ${\alpha\,(Z\alpha)^5}$ correction.
According to Schwinger's textbook~\cite{Sc1970}, 
one can obtain the vacuum-polarization correction
of order $\alpha \, (Z\alpha)^5$ to the Lamb shift by considering the 
slope of the bound-state wave function at the origin.
This holds equally well for the $g$ factor. 
The reason is that the bound-state wave function decays
exponentially as $\exp(-Z \alpha m r)$ whereas the 
Uehling potential decays much faster, namely according to 
Eq.~\eqref{uehl} as $\exp(-\lambda r)$ where $\lambda$
is of the order of the electron rest mass. In the resulting 
product
\begin{align}
|\psi(\vec{r})|^2 \, V_U(r) \sim & \;
\exp( -Z \alpha m r - \lambda r ) 
\nonumber\\[2ex]
= & \; \exp( - \lambda r )\, \left( 1 - Z \alpha m r + 
{\mathcal O}(r^2) \right),
\end{align}
one can thus expand in the first argument of the exponential,
using $\lambda \gg Z \alpha m$. 
The correction term $1 - Z \alpha m r$ 
corresponds to the slope of the wave function at the origin.
A straightforward evaluation gives the following 
vacuum-polarization correction for the ground state:
\begin{equation}
\label{g1VP}
\delta g^{(1)}_{VP} = \frac{\alpha}{\pi} \, (Z\alpha)^4 \,
\left( -\frac{16}{15} + \frac{5 \pi}{9} (Z\alpha) \right) \,,
\end{equation}
which includes the correction of relative order $Z\alpha$.
We here confirm the result of Ref.~\cite{KaIvSh2005}.
For completeness, it is useful to recall the 
corresponding one-loop 
vacuum-polarization correction to the Lamb shift, which 
reads~\cite{Sc1970}
\begin{equation}
\label{E1VP}
\delta E^{(1)}_{VP} = \frac{\alpha}{\pi} \, (Z\alpha)^4 \, m \,
\left( -\frac{4}{15} + \frac{5 \pi}{48} (Z\alpha) \right) \,,
\end{equation}
This concludes the consideration of the one-loop
vacuum-polarization
correction of order $\alpha \, (Z\alpha)^5$ to the $g$ factor.

%
%
\begin{center}
\begin{figure}[htb!]%
\begin{center}
\includegraphics[width=0.95\linewidth]{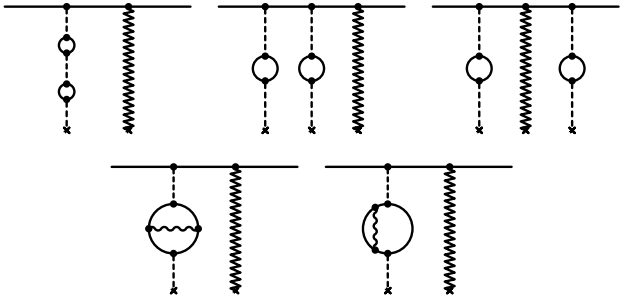}
\end{center}
\caption{\label{fig1} Feynman diagrams for the two-loop vacuum-polarization
corrections to the bound-electron $g$ factor. The first of these is the
loop-after-loop Uehling vacuum-polarization correction and gives a contribution
of $-368 \pi/1701$ in units of $(\alpha/\pi)^2 \, (Z\alpha)^5$ for the $g$
factor. The sum of the second and third diagram (upper row), which are iterated
one-loop perturbations, yields a contribution of $-851 \pi/6804$ in the same
units.  The two last two diagrams (lower row) are K\"{a}ll\'{e}n--Sabry
diagrams.  They lead to a $g$ factor correction of $\pi \, (125176/19845 + 832
\,\ln 2 / 189 - 400 \pi/189)$.}
\end{figure}
\end{center}

%
%
\section{TWO--LOOP CORRECTION}
\label{twoloop}

We now discuss the two-loop calculation. The leading-order 
result~\cite{PaCzJeYe2005} follows as we replace 
the Uehling potential $V_U$ in 
Eq.~\eqref{deltag} by the 
K\"{a}ll\'{e}n--Sabry~\cite{KaSa1955} term.
It reads
\begin{equation}
\delta g_{KS} = \left( \frac{\alpha}{\pi} \right)^2 \,
(Z\alpha)^4 \, \left( - \frac{328}{81} \right)\,.
\end{equation}
The K\"{a}ll\'{e}n--Sabry~term is a genuine two-loop
effect, and one might wonder why the iterated one-loop
diagrams do not also contribute in the order 
$\alpha^2 (Z\alpha)^4$ (these correspond to the 
second and third diagrams in Fig.~\ref{fig1}). 
However, the first-order correction
to the wave function induced by the leading-order magnetic 
interaction $-e \vec{\sigma}\cdot \vec{B}/(2 m)$ vanishes,
and the remaining contribution due to the iterated Uehling
term cancels explicitly in third-order perturbation theory,
because the term with the magnetic interaction 
``in the middle'' cancels against the derivative
term obtained by considering the derivative of the second-order 
Uehling correction with respect to the bound-state energy.

For some of the diagrams in Fig.~\ref{fig1}, 
the $g$ factor correction of order $\alpha^2 (Z\alpha)^5$ can be
obtained by expanding the wave function about the origin,
as it was done for the one-loop theory. 
An example is the first diagram in Fig.~\ref{fig1},
which can be expressed as the expectation value of the 
loop-after-loop Uehling potential, evaluated with 
wave functions perturbed by the magnetic interaction.
For the other diagrams,
the calculation is more complicated. In particular, since the 
iterated Uehling correction 
(second and third diagram in the first row 
in Fig.~\ref{fig1}) contributes at the order 
of $\alpha^2 (Z\alpha)^5$, one cannot avoid the complete
calculation
of the (first-order) perturbation to the wave function 
by the Uehling correction, which involves 
exponentials, exponential integrals, 
logarithms and powers of the radial variable. 
E.g., the magnetic interaction in the middle vertex 
demands a further integration over the radial coordinate.
The further calculation 
proceeds along the lines outlined in Ref.~\cite{Pa1993pra}
for the two-loop vacuum-polarization 
corrections to the Lamb shift.

We finally obtain for the two-loop binding
contribution $\delta g^{(2)}_{VP}$ due to the 
diagrams in Fig.~\ref{fig1},
\begin{align}
\label{g2VP}
\delta g^{(2)}_{VP} =& \; \left( \frac{\alpha}{\pi} \right)^2 \,
(Z\alpha)^4 \, \left[ - \frac{328}{81} +
(Z\alpha) \, \pi \, \left( \frac{1420807}{238140}
\right. \right.
\nonumber\\[2ex]
& \; \left. \left. +
\frac{832}{189} \ln 2 -
\frac{400}{189} \pi \right) \right] 
\nonumber\\[2ex]
= & \; \left( \frac{\alpha}{\pi} \right)^2 \, (Z\alpha)^4 \, 
\left[ -4.049 + 7.442\, (Z\alpha) \right] \,.
\end{align}
The numerical coefficient of the $(Z\alpha)$-correction is 
rather large, mainly because it has a factor $\pi$ 
in the numerator. 

Just as for the one-loop calculation, it is useful 
to compare our results to those for the Lamb shift,
selecting the corresponding
set of diagrams. For the Lamb shift, we can identify 
the diagrams corresponding to those in Fig.~\ref{fig1} 
by simply eliminating the interaction with the external 
magnetic field. The resulting diagrams after this removal 
operation are equivalent to 
the diagrams labeled as IV and VI in Ref.~\cite{Pa1993pra}.
The corresponding contribution to the Lamb-shift 
is~\cite{EiGrOw1992,Pa1993pra}
\begin{align}
\label{E2VP}
\delta E^{(2)}_{VP} =& \; \left( \frac{\alpha}{\pi} \right)^2 \, m \,
(Z\alpha)^4 \, \left[ - \frac{82}{81} +
(Z\alpha) \, \pi \, \left( \frac{7421}{6615} 
\right. \right.
\nonumber\\[2ex]
& \; \left. \left. + \frac{52}{63} \ln 2 -
\frac{25}{63} \pi \right) \right] \nonumber\\[2ex]
=& \;  \left( \frac{\alpha}{\pi} \right)^2 \, (Z\alpha)^4 \, 
\left[ -1.012 + 1.405\, (Z\alpha) \right] \,,
\end{align}
and we have verified it using our approach.
This concludes our two-loop vacuum-polarization calculations.

%
%
\section{SUMMARY}
\label{conclusions}

In this brief report, we describe the evaluation
of a part of the binding, vacuum-polarization correction to
the bound-electron $g$ factor.
The vacuum-polarization corrections represent 
a preparatory calculation for the self-energy corrections,
which are much more difficult to evaluate.
In view of the multitude of terms generated in comparison
to the corresponding self-energy correction to the Lamb shift of
order $\alpha (Z\alpha)^5$, and in view of the additional
complexity of the calculation due to the added external 
magnetic field, considerable difficulties are expected.

It may, already at this point, be permitted to speculate a
little about the magnitude of the complete correction
to the $g$ factor of order $\alpha^2 (Z\alpha)^5$,
which is less than an estimate but perhaps more than 
just guesswork. 
Namely, we observe there appears to be a rather universal factor 
in the range of $3.5 \ldots 5.5$ by which the $g$ factor
coefficients of a given order in the $Z\alpha$-expansion
are larger than the corresponding Lamb shift coefficients
for the ground state. In particular, we compare in the 
order $\alpha(Z\alpha)^4$, the 
coefficient $-16/15$ in \eqref{g1VP} to 
the coefficient $-4/15$ in Eq.~\eqref{E1VP} 
(the $g$ factor coefficient is larger than the 
Lamb shift coefficient by a relative factor four).
At relative order $Z\alpha$, the relative factor is 
$5.33$ (the coefficients are $5\pi/9$ versus~$5\pi/48$).
For the $g$ factor at two-loop order, the relative 
factor at $\alpha^2 (Z\alpha)^5$ is $5.30$,
as evident from Eqs.~\eqref{g2VP} and~\eqref{E2VP}.
A factor in the range $3.5 \dots 5.5$ also appears for the 
self-energy corrections.
We recall that the complete two-loop correction
to the Lamb shift in the order 
$\alpha^2 (Z\alpha)^5$ is~\cite{Pa1993,Pa1994prl,EiSh1995,EiGrSh1997} 
\begin{equation}
\delta E^{(2)}_5 = \; -21.55 \, 
\left( \frac{\alpha}{\pi} \right)^2 \, 
(Z\alpha)^5 \, m \,.
\end{equation}
Our ``educated guess'' for the complete correction 
to the $g$ factor thus is
\begin{equation}
\delta g^{(2)}_5 = \; C \,
\left( \frac{\alpha}{\pi} \right)^2 \, 
(Z\alpha)^5 \, \,, \quad
-118 < C < -75 \,.
\end{equation}
The magnitude of this estimate of the coefficient generates obvious 
interest.

%
%
\section*{Acknowledgments}

The authors acknowledges helpful discussions with K. Pachucki and 
E. Remiddi, and helpful remarks by an anonymous referee.

\end{document}